\documentclass{ws-ijmpd}
\begin{document} \title{Observational constraints on dark energy
model} \author{Yungui Gong}
\address{College of Electronic
Engineering, Chongqing University of Posts and Telecommunications,
Chongqing 400065, P.R. China\\gongyg@cqupt.edu.cn}
\maketitle
\begin{abstract}
The recent observations support that our Universe
is flat and expanding with acceleration. We analyze a general
class of quintessence models by using the recent type Ia supernova
and the first year Wilkinson Microwave Anisotropy Probe (WMAP)
observations. For a flat universe dominated by a dark energy with
constant $\omega$ which is a special case of the general model, we
find that $\Omega_{\rm m0}=0.30^{+0.06}_{-0.08}$ and $\omega_{\rm
Q}\le -0.82$, and the turnaround redshift $z_{\rm T}$ when the
universe switched from the deceleration phase to the acceleration
phase is $z_{\rm T}=0.65$. For the general model, we find that
$\Omega_{\rm m0}\sim 0.3$, $\omega_{\rm Q0}\sim -1.0$, $\beta\sim
0.5$ and $z_{\rm T}\sim 0.67$. A model independent polynomial
parameterization of dark energy is also considered, the best fit
model gives $\Omega_{\rm m0}=0.40\pm 0.14$, $\omega_{\rm Q0}=-1.4$
and $z_{\rm T}=0.37$. \end{abstract}

\section{Introduction}
The type Ia supernova (SN Ia) observations indicate that the
expansion of the Universe is speeding up rather than slowing down
\cite{sp97}-\cite{riess}. The measurement of the anisotropy of the
cosmic microwave background (CMB) favors a flat universe
\cite{pdb00,bennett03,DNSpergel}. The observation of type Ia
supernova SN 1997ff at $z\sim 1.7$ also provides the evidence that
the Universe is in the acceleration phase and was in the
deceleration phase in the past \cite{riess,agr}. The transition
from the deceleration phase to the acceleration phase happened
around the redshift $z_{\rm T}\sim 0.4$ \cite{riess,mstagr}. In
this paper, we use the notation $z_{\rm T}$ for the transition
redshift. A new component with negative pressure widely referred
as dark energy is usually introduced to explain the accelerating
expansion. The simplest form of dark energy is the cosmological
constant with the equation of state parameter $\omega_{\rm
\Lambda}=-1$. One easily generalizes the cosmological constant
model to dynamical cosmological constant models such as the dark
energy model with negative constant equation of state parameter
$-1\le \omega_{\rm Q}<-1/3$ and the holographic dark energy models
\cite{holo}. If we remove the null energy condition restriction
$\omega_{\rm Q}\ge -1$ to allow supernegative $\omega_{\rm Q}<-1$,
then we have the phantom energy models \cite{phantom}. More exotic
equation of state is also possible, such as the Chaplygin gas
model with the equation of state $p=-A/\rho$ and the generalized
Chaplygin gas model with the equation of state $p=-A/\rho^\alpha$
\cite{chaply}. In general, a scalar field $Q$ that slowly evolves
down its potential $V(Q)$ takes the role of a dynamical
cosmological constant. The scalar field $Q$ is also called the
quintessence field \cite{quint}-\cite{gong02}. The energy density
of the quintessence field must remain very small compared with
that of radiation or matter at early epoches and evolves in a way
that it started to dominate the universe around the redshift
$0.4$. Instead of the quintessence field with the usual kinetic
term $\dot{Q}^2/2$, tachyon field as dark energy was also proposed
\cite{tachyon}. The tachyon models have the accelerated phase
followed by the decelerated phase.

Although most dark energy models are consistent with current
observations, the nature of dark energy is still mysterious.
Therefore it is also possible that the observations show a sign of
the breakdown of the standard cosmology. Some alternative models
to dark energy models were proposed along this line of reasoning.
These models are motivated by extra dimensions. In these models,
the usual Friedmann equation $H^2=8\pi G\rho/3$ is modified to a
general form $H^2=g(\rho)$ and the universe is composed of the
ordinary matter only \cite{freese02}-\cite{gong03}. In other
words, the dark energy component is unnecessary.

In this paper, I first use the 58 SN Ia data in
Ref.~\refcite{raknop03}, the 186 SN Ia data in
Ref.~\refcite{riess} and WMAP data \cite{DNSpergel} to constrain
the parameter space of a general class of quintessence models
discussed in Ref.~\refcite{gong02}. In that model, a general
relation between the potential energy and the kinetic energy of
the quintessence field was proposed. As we know, the average
kinetic energy is the same as the average potential energy for a
point mass in a harmonic oscillator. For a stable,
self-gravitating, spherical distribution of equal mass objects,
the total kinetic energy of the objects is equal to minus 1/2
times the total gravitational potential energy. Therefore, the
physics of dark energy may be determined if the relationship
between the potential energy and the kinetic energy is known. Then
I consider three different model independent parameterizations of
$\omega_{\rm Q}$ to find out some properties of dark energy. After
we determine the parameters in these models, the transition
redshift $z_{\rm T}$ is obtained. The paper is organized as
follows. After a brief introduction in section 1, the general
class of models is reviewed in section 2. In section 3, I discuss
the methodology used in this paper. In section 4, I give the main
fitting results. In section 5, I conclude the paper by using a
model independent analysis and compare the results with those in
the literature.

\section{Model Review}
For a spatially flat, isotropic and homogeneous universe with both
an ordinary pressureless dust matter and a minimally coupled
scalar field $Q$ source, the Friedmann equations are
\begin{eqnarray}
\label{cos1} H^2\equiv\left({\dot{a}\over a}\right)^2={8\pi G\over
3}(\rho_{\rm m}+\rho_{\rm Q}),\\ \label{cos2} {\ddot{a}\over
a}=-{4\pi G\over 3}(\rho_{\rm m}+\rho_{\rm Q}+3p_{\rm Q}),\\
\label{cos3} \ddot{Q}+3H\dot{Q}+V'(Q)=0,
\end{eqnarray}
where dot means derivative with respect to time, $\rho_{\rm
m}=\rho_{\rm m0}(a_0/a)^3$ is the matter energy density, a
subscript 0 means the value of the variable at present time,
$\rho_{\rm Q}=\dot{Q}^2/2+V(Q)$, $p_{\rm Q}=\dot{Q}^2/2-V(Q)$,
$V'(Q)=dV(Q)/dQ$ and $V(Q)$ is the potential of the quintessence
field. In Ref.~\refcite{gong02}, a general relationship
\begin{equation}
\label{assm1}
V(Q)=\beta \dot{Q}^2+C,
\end{equation}
was proposed instead of assuming a particular potential for the
quintessence field or a particular form of the scale factor, where
$\beta$ and $C$ are constants. Note that the above equation
(\ref{assm1}) is a constraint equation, one should not just
substitute the above equation into the Lagrangian and thinks that
the model is equivalent to a $1/2+\beta$ kinetic term plus a
cosmological constant term $C$. The above general potential
includes the hyperbolic potential and the double exponential
potential. In terms of $\rho_{\rm Q0}$ and $\omega_{\rm Q0}$, we
have
\begin{eqnarray}
C&=&\left[{1\over 2}-\beta-\left({1\over
2}+\beta\right)\omega_{\rm Q0}\right]\rho_{\rm Q0},\\ \label{eng1}
\rho_{\rm Q}&=&(1/2+\beta)(1+\omega_{\rm Q0})\rho_{\rm
Q0}\left({a_0\over a}\right)^{6/(2\beta+1)} +\left[{1\over
2}-\beta-\left({1\over 2}+\beta\right)\omega_{\rm
Q0}\right]\rho_{\rm Q0},\\ \label{qup1} p_{\rm
Q}&=&(1/2-\beta)(1+\omega_{\rm Q0})\rho_{\rm Q0}\left({a_0\over
a}\right)^{6/(2\beta+1)} -\left[{1\over 2}-\beta-\left({1\over
2}+\beta\right)\omega_{\rm Q0}\right]\rho_{\rm Q0},\\ \label{hub}
H^2&=&{8\pi G\over 3}\left\{\rho_{\rm m0}\left(a_0\over
a\right)^3+(1/2+\beta)(1+\omega_{\rm Q0})\rho_{\rm
Q0}\left({a_0\over a}\right)^{6/(2\beta+1)} +C\right\}.
\end{eqnarray}
To make the quintessence field sub-dominated during early times,
we require that $\beta\ge 0.5$. The transition from deceleration
to acceleration happens when the deceleration parameter
$q=-\ddot{a}H^2/a=0$. From equations (\ref{cos2}), (\ref{eng1})
and (\ref{qup1}), in terms of the redshift parameter $1+z=a_0/a$,
we have
\begin{equation}
\label{rel3}  (1+z_{\rm T})^3+2(1-\beta)(1+\omega_{\rm
Q0}){\rho_{\rm Q0}\over \rho_{\rm m0}}(1+z_{\rm
T})^{6/(2\beta+1)}-[1-2\beta-(1+2\beta)\omega_{\rm Q0}]{\rho_{\rm
Q0}\over \rho_{\rm m0}}=0.
\end{equation}
This equation gives a relationship between $\omega_{\rm Q0}$ and
$\Omega_{\rm Q0}$. Now let us turn our attention to two special
cases.

Case 1: $C=0$, the equation of state of the scalar field is a
constant, $\omega_{\rm Q}=(1/2-\beta)/(1/2+\beta)$. The potential
is \cite{johri,lautm}
$$V(Q)=A[\sinh k(Q/\alpha+B)]^{-\alpha},$$
where $\alpha=2/(\beta-1/2)$, $k^2=48\pi G/(2\beta+1)$,
$A^{\beta-1/2}=(1/2+\beta)C_2^{2\beta+1}\beta^{\beta-1/2}/(\rho_{\rm
m0}a^3_0)$ and $B$ is an arbitrary integration constant.

Case 2: $\beta=1/2$, the pressure of the scalar field becomes a
constant $p_{\rm Q}=-C=\omega_{\rm Q0}\rho_{\rm Q0}$ and the
potential is the double exponential potential \cite{aasss}
$$V(Q)={A^2\over 8}[\exp(2\alpha Q)+\exp(-2\alpha Q)]+{\alpha^2A^2\over 6\pi G}-{A^2\over 4}.$$
The constant pressure model is equivalent to an ordinary matter
with effective matter content $\Omega^{\rm eff}_{\rm
m0}=(1+\omega_{\rm Q0})\Omega_{\rm Q0}+ \Omega_{\rm m0}$ plus a
cosmological constant $\rho_\Lambda=-\omega_{\rm Q0}\rho_{\rm
Q0}$.
\section{Methodology}
In order to use the WMAP result, one usually parameterizes the
location of the $m$-th peak of CMB power spectrum as \cite{doran}
$$l_m=(m-\phi_m)l_{\rm A},$$
where the acoustic scale $l_{\rm A}$ is
\begin{equation}
l_{\rm A}={\pi\over \bar{c_{\rm s}}}{\tau_0-\tau_{\rm ls}\over
\tau_{\rm ls}}={\pi\over \bar{c_{\rm s}}}{\int^{z_{\rm ls}}_0  d
z/\sqrt{g(z)}\over \int^{\infty}_{z_{\rm ls}}  d z/\sqrt{g(z)}},
\end{equation}
the conformal time at the last scattering $\tau_{\rm ls}$ and at
today $\tau_0$ are
\begin{eqnarray}
\tau_{\rm ls}&=&\int^{\tau_{\rm ls}}_0
 d\tau=\int^{\infty}_{z_{\rm ls}} { d z\over a_0
H_0\sqrt{g(z)}},\\ \tau_0&=&\int^{\tau_0}_0  d\tau=\int^{\infty}_0
{ d z\over a_0 H_0\sqrt{g(z)}},\\   g(z)&=&\Omega_{\rm m0}(1+z)^3
+\Omega_{\rm r0}(1+z)^4+(1/2+\beta)
(1+\omega_{\rm Q0})\Omega_{\rm Q0}(1 +z)^{6/(2\beta+1)}\nonumber\\
&&+\left[{1\over 2}-\beta-\left({1\over 2}+\beta\right)\omega_{\rm
Q0}\right]\Omega_{\rm Q0},
\end{eqnarray}
$\Omega_{\rm r0}=8.35\times 10^{-5}$ is the current radiation
component and $z_{\rm ls}=1089\pm 1$ \cite{bennett03}. The
difficulty of this method is that there are several undetermined
parameters, such as $\phi_m$ and $\bar{c}_s$. Instead, we use the
CMB shift parameter $\mathcal{R}\equiv \Omega_{\rm
m0}^{1/2}a_0H_0(\tau_0-\tau_{\rm ls})=1.716\pm 0.062$ \cite{wang1}
to constrain the model.

The luminosity distance $d_{\rm L}$ is defined as
\begin{eqnarray}
\label{lumin} d_{\rm L}(z)=a_0c(1+z)\int^{t_0}_{\rm t} { d t'\over
a(t')}=c{1+z\over H_0}\int^z_0  d u[g(u)]^{-1/2} .
\end{eqnarray}
The apparent magnitude redshift relation becomes
\begin{eqnarray}
\label{magn} m(z)&=&M+5\log_{10}d_{\rm L}(z)
+25=\mathcal{M}+5\log_{10}\mathcal{D}_{\rm L}(z)\nonumber\\
&=&\mathcal{M}+5\log_{10}\left[c(1+z)\int^z_0  d u
[g(u)]^{-1/2}\right],
\end{eqnarray}
where $\mathcal{D}_{\rm L}(z)=H_0d_{\rm L}(z)$ is the
"Hubble-constant-free" luminosity distance, $M$ is the absolute
peak magnitude and $\mathcal{M}=M-5\log_{10}H_0+25$. $\mathcal{M}$
can be determined from the low redshift limit at where
$\mathcal{D}_{\rm L}(z)=z$. We use the 54 SNe Ia data with both
the stretch correction and the host-galaxy extinction correction,
i.e., the fit 3 supernova data in Ref.~\refcite{raknop03} (we
refer the data as Knop sample), and the 186 SNe Ia data in
Ref.~\refcite{riess} (we refer the data as Riess sample) to
constrain the model. The parameters in the model are determined
using a $\chi^2$-minimization procedure based on MINUIT code.
There are four parameters in the fit: the current mass density
$\Omega_{\rm m0}$, the current dark energy equation of state
parameter $\omega_{\rm Q0}$, the constant $\beta$ as well as the
nuisance parameter $\mathcal{M}$. The range of parameter space is
$\Omega_{\rm m0}=[0,\ 1]$ and $\omega_{\rm Q0}=(-1,\ 0]$.
\section{Results}
\begin{figure}[htb]
\vspace{-0.1in}
\begin{center}
$\begin{array}{cc} \epsfxsize=2.5in \epsffile{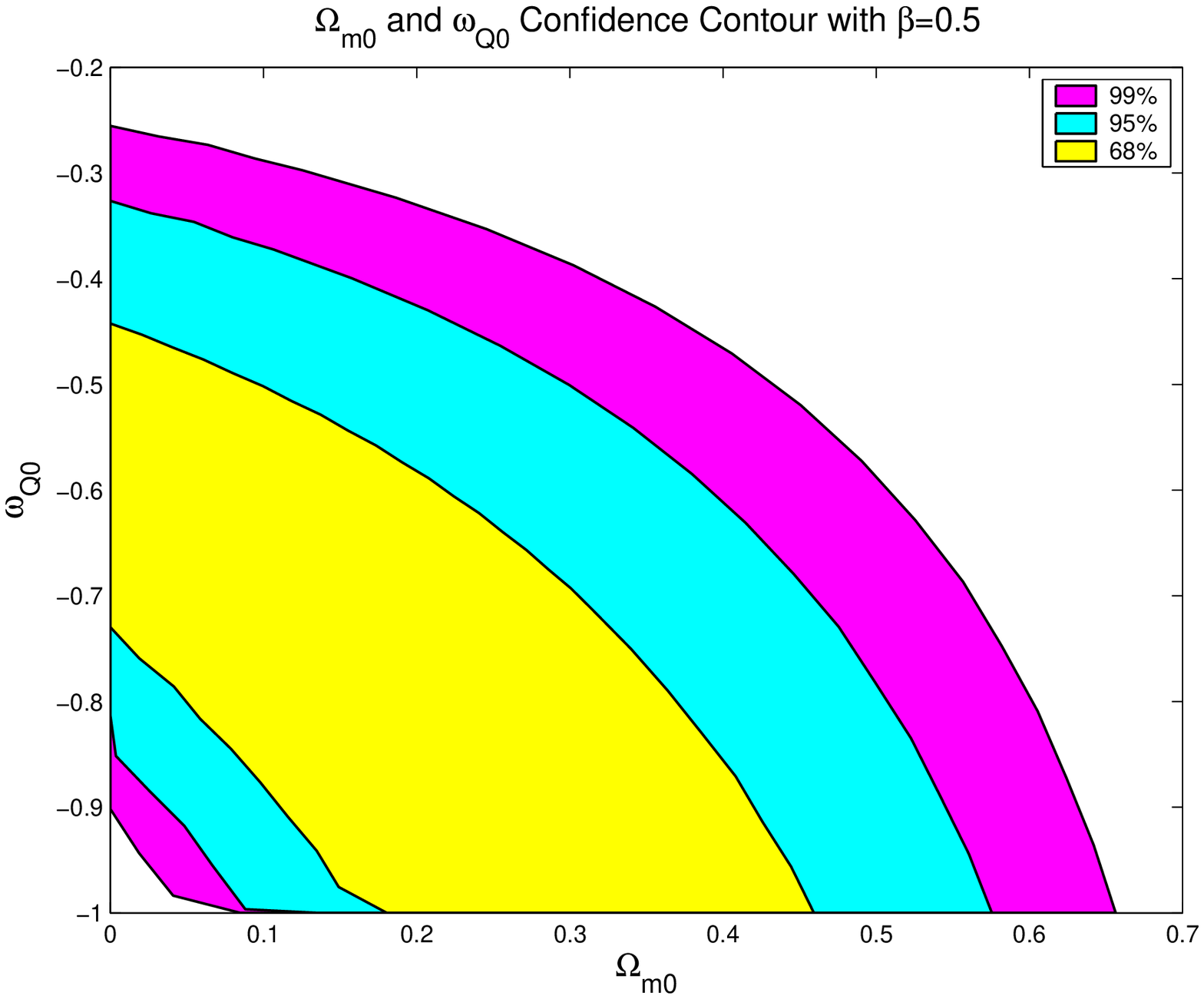}&
\epsfxsize=2.5in \epsffile{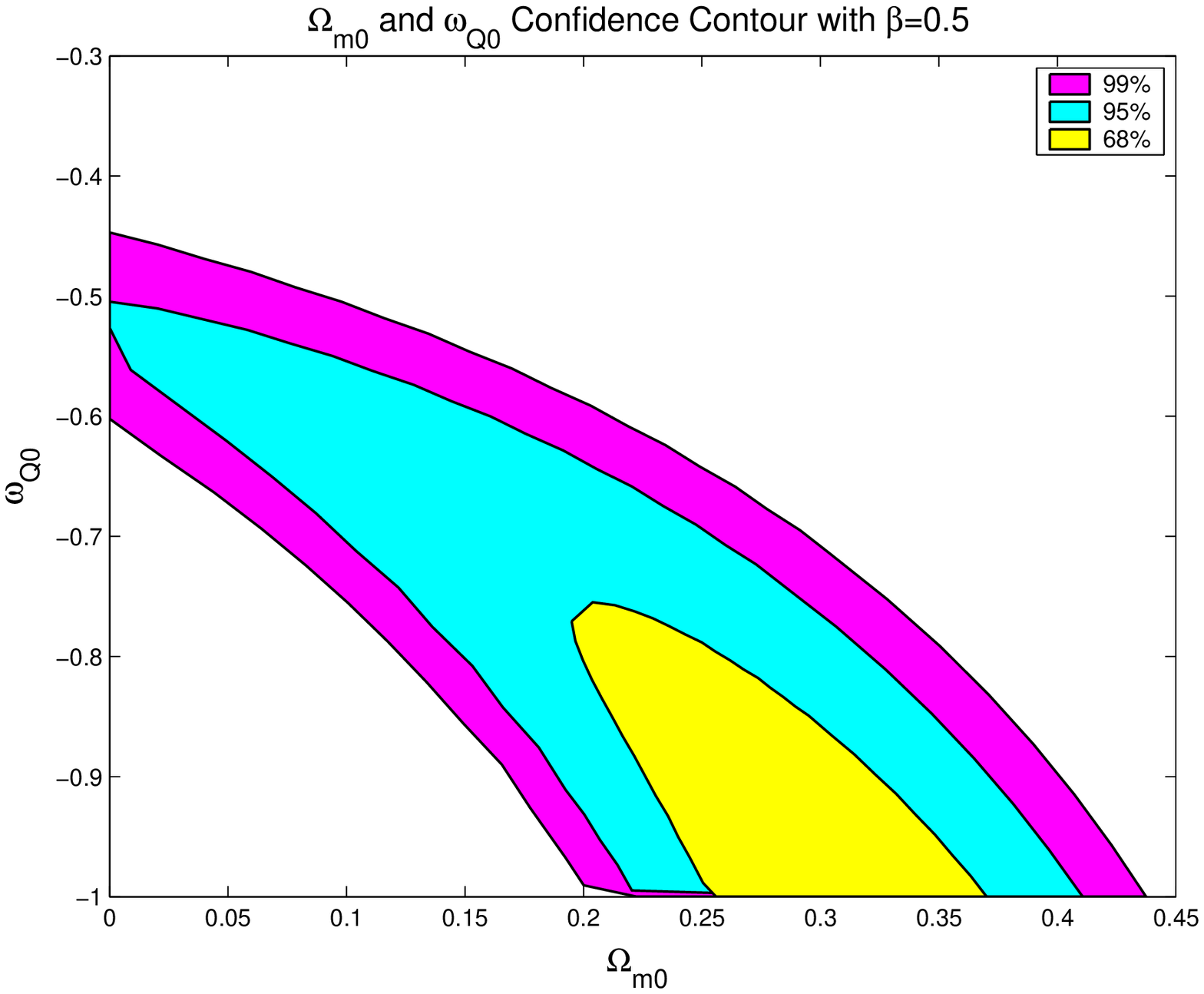}\\
\epsfxsize=2.5in \epsffile{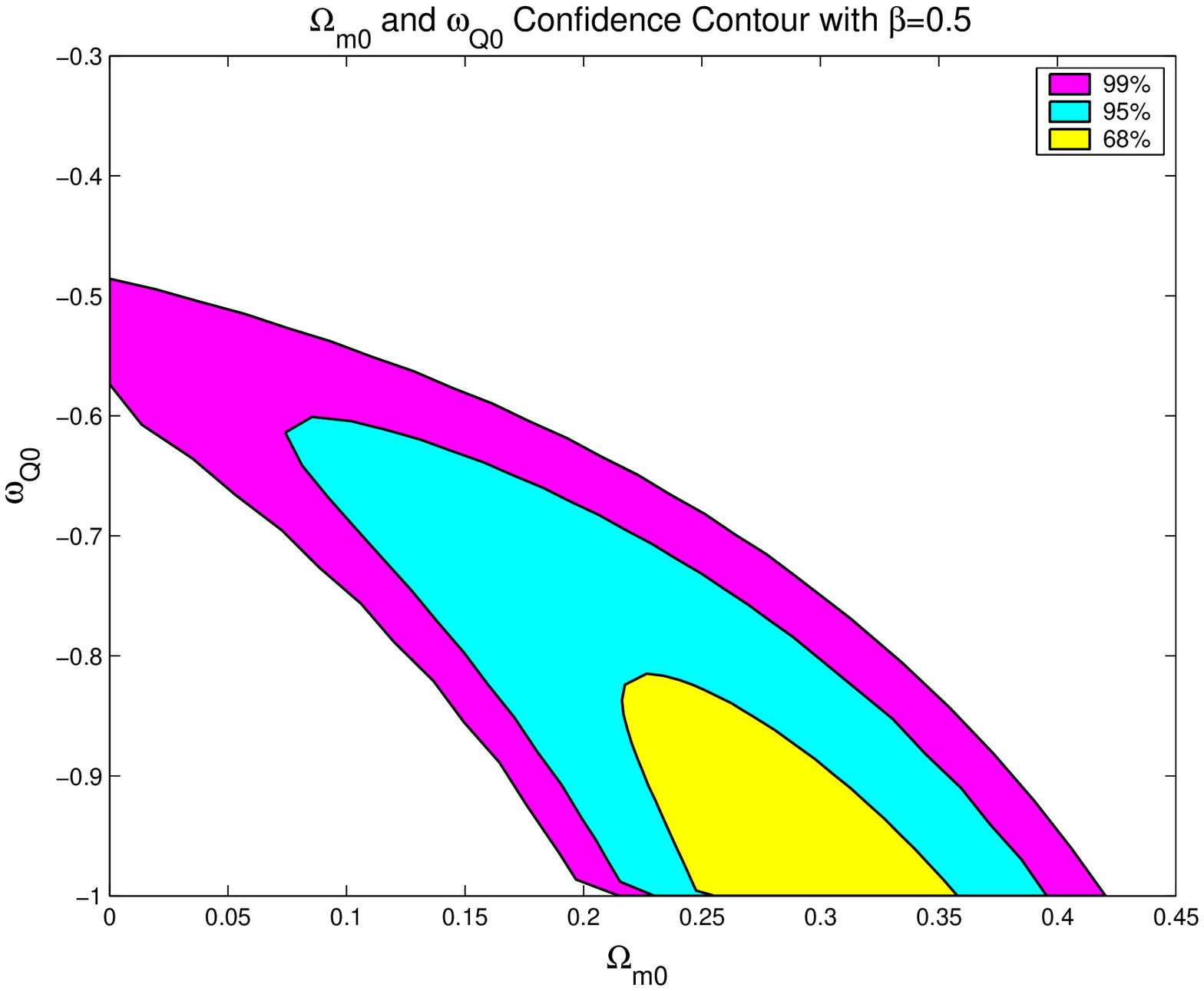}& \epsfxsize=2.5in
\epsffile{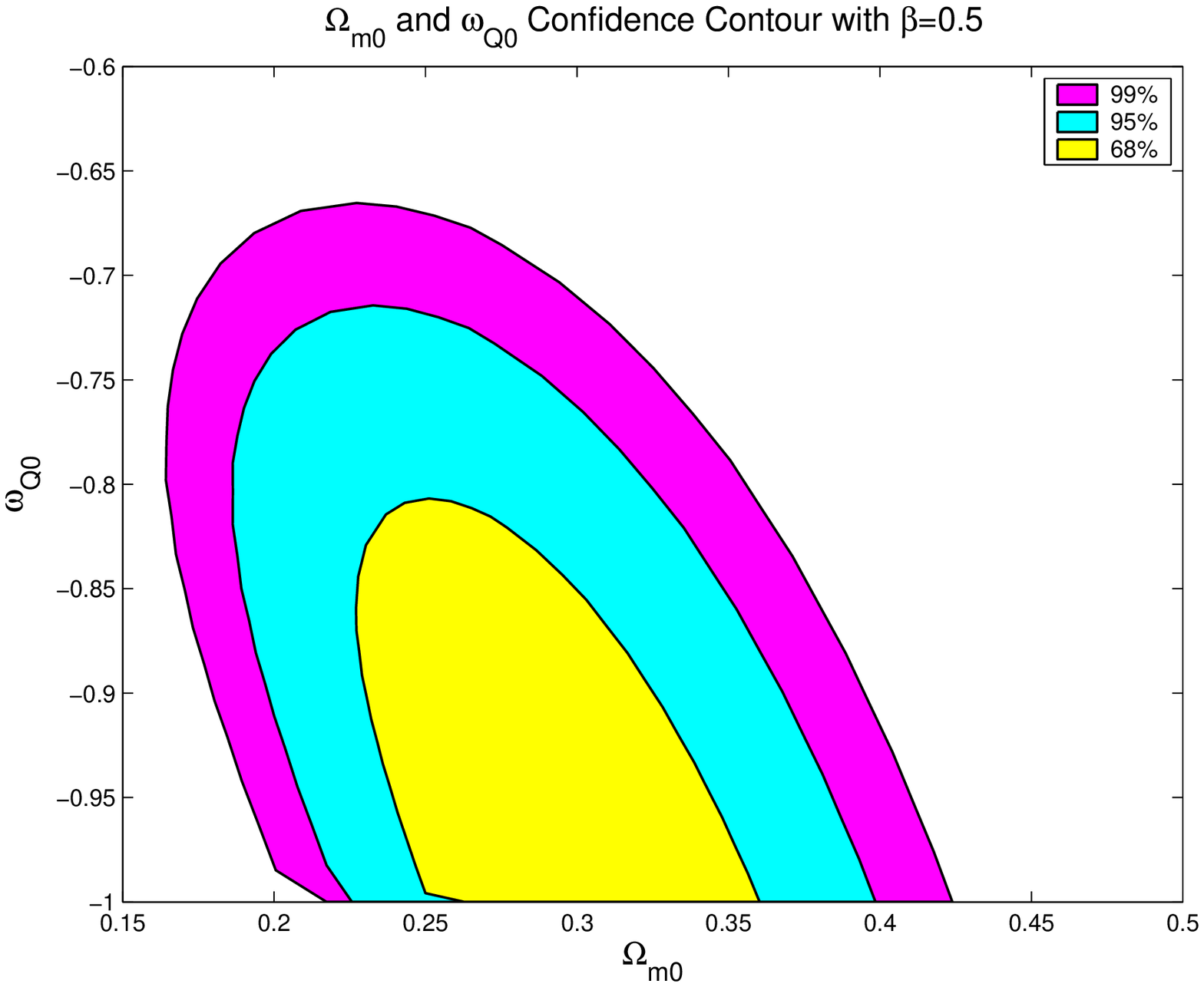}
\end{array}$
\end{center}
\vspace{-0.2in} \caption{The confidence contours of $\Omega_{\rm
m0}$ and $\omega_{\rm Q}$ for the $C=0$ model. The left upper
panel is the 68\%, 95\% and 99\% confidence regions fitted from
Knop sample. The right upper panel is the 68\%, 95\%, 99\%
confidence regions fitted from Riess gold sample, the left lower
panel is the 68\%, 95\% and 99\% confidence regions fitted from
Riess gold and silver sample. The right lower panel is the 68\%,
95\%, 99\% confidence regions fitted from Riess gold sample and
WMAP data.} \label{consfit}
\end{figure}

For the dark energy model with constant $\omega_{\rm Q}$, i.e.,
the model with $C=0$, the best fit parameters to the 54 knop
sample are $\Omega_{\rm m0}=[0,\ 0.46]$ centered at 0.30 and
$\omega_{\rm Q}=(-1.0,\ -0.44]$ centered at almost $-1.0$ with
$\chi^2=45.6$ at 68\% confidence level. The best fit parameters to
the 157 Riess gold sample are $\Omega_{\rm m0}=[0.19,\ 0.37]$
centered at 0.31 and $\omega_{\rm Q}=(-1,\ -0.75]$ centered at
almost $-1.0$ with $\chi^2=177.1$ at 68\% confidence level. The
best fit parameters to the 186 Riess gold and silver sample are
$\Omega_{\rm m0}=[0.22,\ 0.36]$ centered at 0.30 and $\omega_{\rm
Q}=(-1,\ -0.81]$ centered at almost $-1.0$ with $\chi^2=232.3$ at
68\% confidence level. The best fit parameters to the Riess gold
sample and WMAP data combined are $\Omega_{\rm m0}=[0.22,\ 0.36]$
centered at 0.30 and $\omega_{\rm Q}=(-1.0,\ -0.81]$ centered at
almost $-1.0$ with $\chi^2=177.3$ at 68\% confidence level. The
best fit parameters to the Riess gold and silver sample and WMAP
data combined are $\Omega_{\rm m0}=[0.23,\ 0.35]$ centered at 0.30
and $\omega_{\rm Q}=(-1.0,\ -0.84]$ centered at almost $-1.0$ with
$\chi^2=232.4$ at 68\% confidence level. From the above results,
it is shown that the addition of Riess silver data give almost the
same results as those from Riess gold sample only. Therefore, we
will use Riess gold sample only in the following discussion. The
confidence regions of $\Omega_{\rm m0}$ and $\omega_{\rm Q}$ are
shown in figure \ref{consfit}.

\begin{figure}[htb]
\vspace{-0.1in}
\begin{center}
$\begin{array}{cc} \epsfxsize=2.5in \epsffile{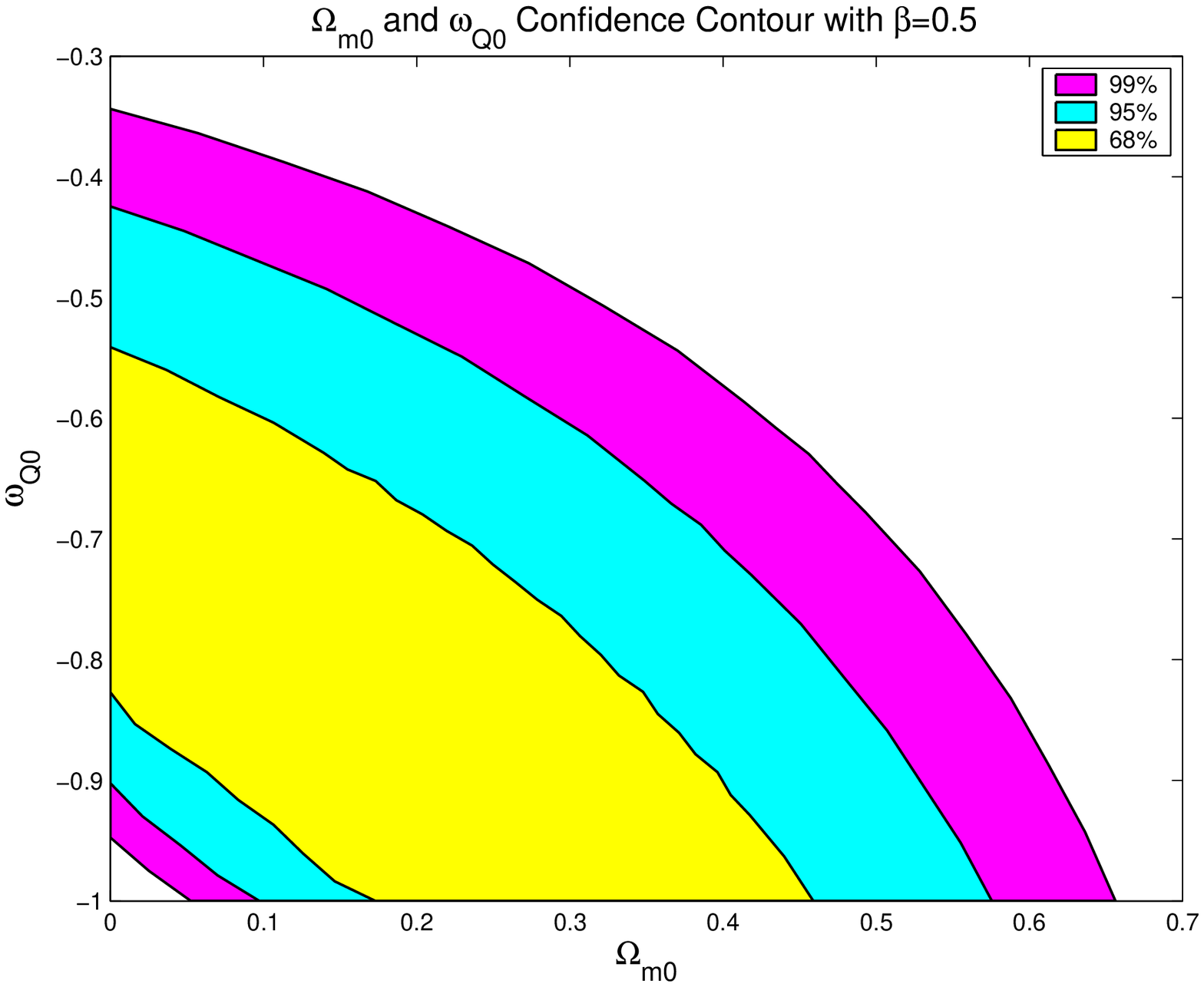}&
\epsfxsize=2.5in \epsffile{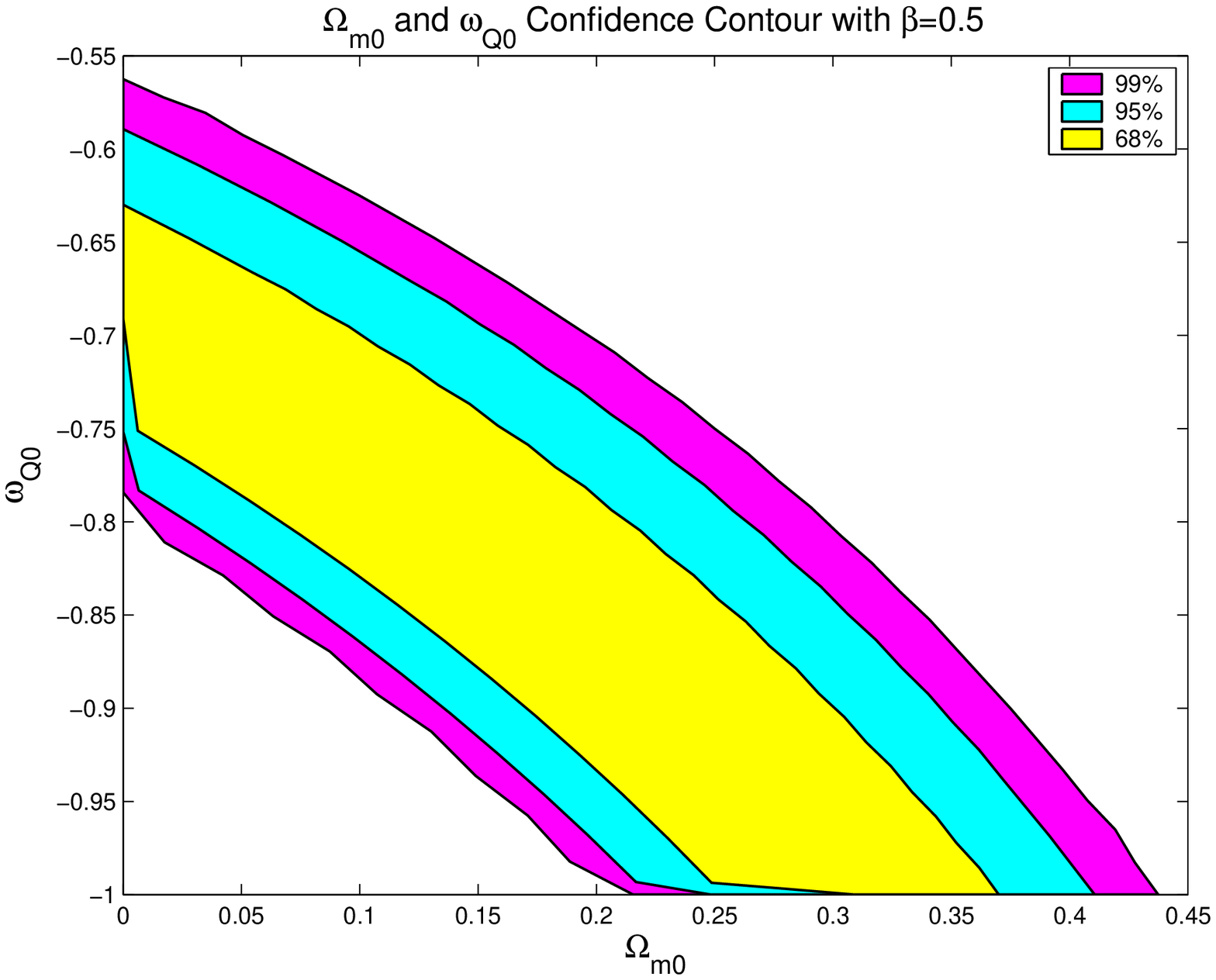}
\end{array}$
\end{center}
\vspace{-0.2in} \caption{The 68\%, 95\% and 99\% confidence
contours of $\Omega_{\rm m0}$ and $\omega_{\rm Q0}$ for
$\beta=1/2$. The left panel shows fits to Knop sample and the
right panel shows fits to Riess gold sample.} \label{specfit}
\end{figure}

\begin{figure}[htb]
\vspace{-0.1in}
\begin{center}
 \epsfxsize=4in \epsffile{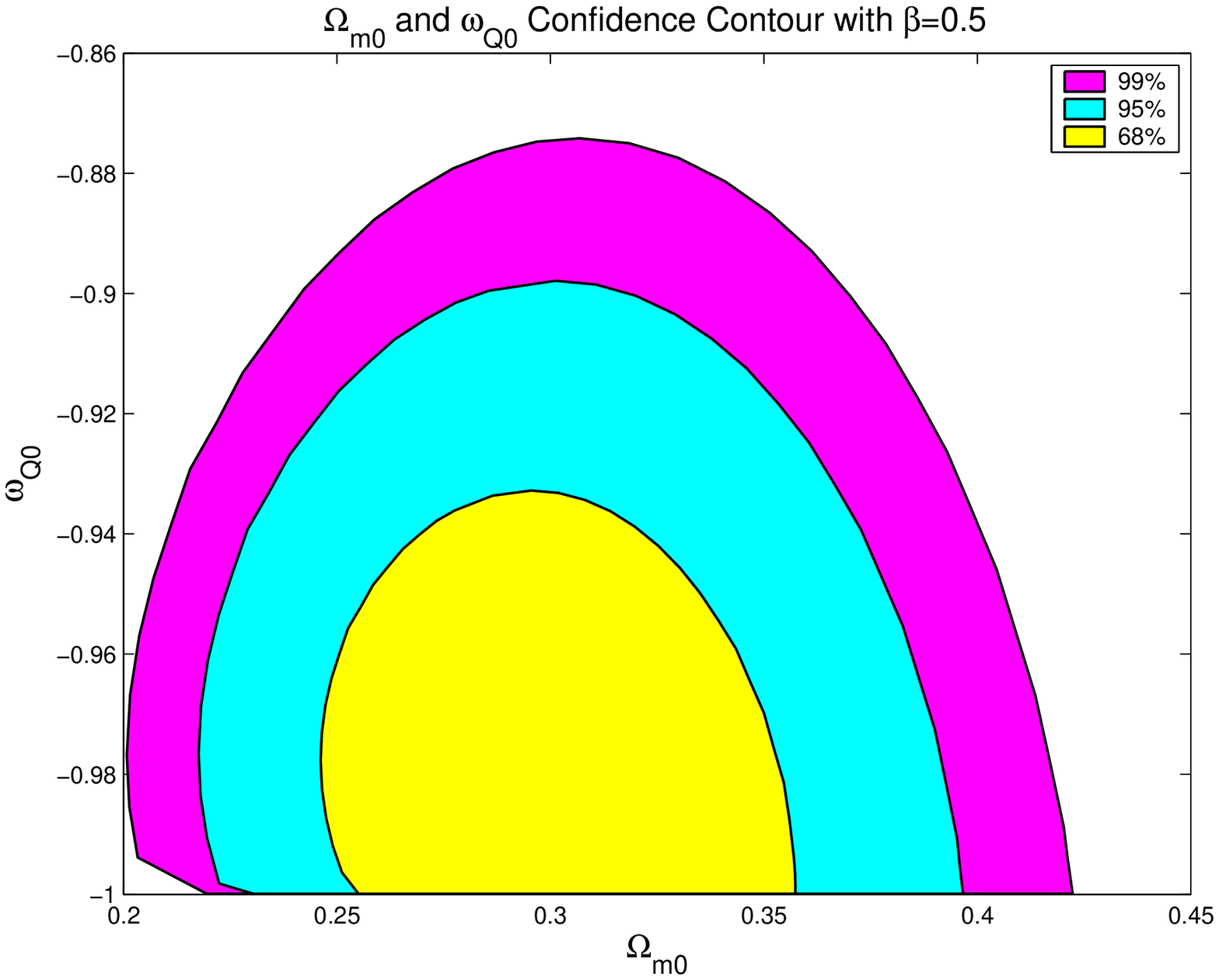}
\end{center}
\vspace{-0.2in} \caption{The 68\%, 95\% and 99\% confidence
contours of $\Omega_{\rm m0}$ and $\omega_{\rm Q0}$ for
$\beta=1/2$ with Riess gold sample and WMAP data. }
\label{specfitrm}
\end{figure}

For the general model $C\neq 0$, we first analyze the special
constant pressure model $\beta=1/2$ which is equivalent to the
$\Lambda$-CDM mdoel. The best fits to the 54 Knop sample are
$\Omega_{\rm m0}=[0,\ 0.46]$ centered at almost zero and
$\omega_{\rm Q0}=(-1,\ -0.54]$ centered at $-0.71$ with
$\chi^2=45.6$ at 68\% confidence level. The best fits to the 157
Riess gold sample are: $0\le \Omega_{\rm m0}\le 0.37$ and $-1<
\omega_{\rm Q0}\le -0.63$ at 68\% confidence level with
$\chi^2=177.1$. Note that the effective $\Omega_{\rm m0}\sim 0.3$
although the best fit $\Omega_{\rm m0}$ is almost zero. The best
fits to the 157 Riess gold sample and WMAP data combined are:
$0.25\le \Omega_{\rm m0}\le 0.35$ and $-1< \omega_{\rm Q0}\le
-0.93$ with $\chi^2=177.1$ at 68\% confidence level. The
confidence regions of $\Omega_{\rm m0}$ and $\omega_{\rm Q0}$ are
shown in figures \ref{specfit} and \ref{specfitrm}.

For the general model, the range of parameter space is
$\Omega_{\rm m0}=[0,\ 1]$, $\omega_{\rm Q0}=(-1,\ 0]$ and
$\beta\ge 0.5$. The best fits to the 54 Knop sample are: $0\le
\Omega_{\rm m0}\le 0.50$, $-1< \omega_{\rm Q0}\le -0.23$ and
$\beta$ varies in a big range with $\chi^2=45.6$ at 68\%
confidence level. The best fits to the 157 Riess gold sample are:
$0\le \Omega_{\rm m0}\le 0.38$, $-1< \omega_{\rm Q0}\le -0.57$ and
$\beta$ varies in a big range with $\chi^2=177.1$ at 68\%
confidence level. The best fits to the 157 Riess gold sample and
WMAP data combined are: $0.21\le \Omega_{\rm m0}\le 0.37$, $-1.0<
\omega_{\rm Q0}\le -0.75$ and $\beta\ge 0.5$ with $\chi^2=177.1$
at 68\% confidence level. From the above results, we see that the
best fit model tends to be the $\Lambda$-CDM model with
$\Omega_{\rm m0}\sim 0.3$. \begin{figure}[htb] \vspace{-0.1in}
\begin{center} $\begin{array}{cc} \epsfxsize=2.5in
\epsffile{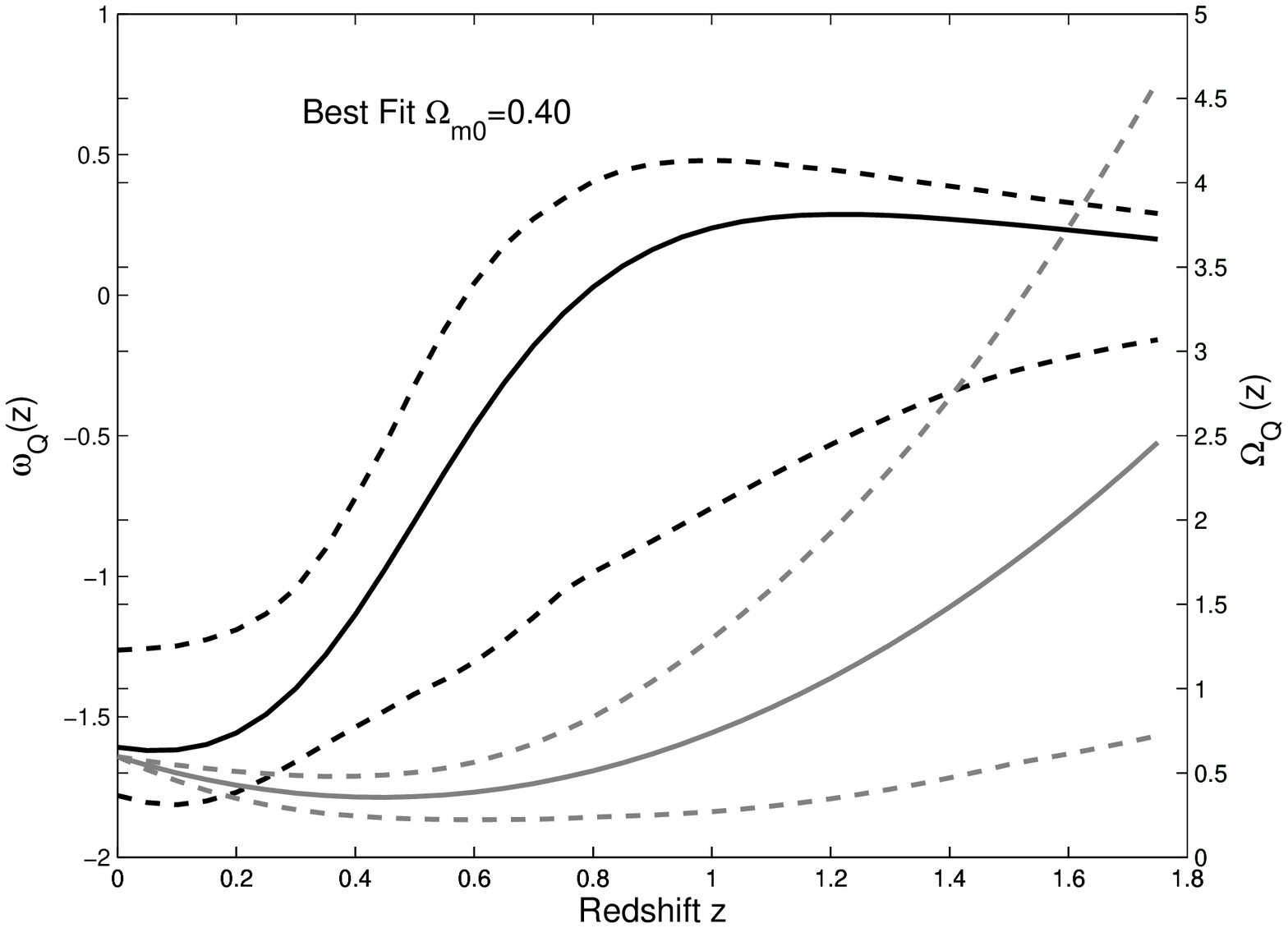}& \epsfxsize=2.5in \epsffile{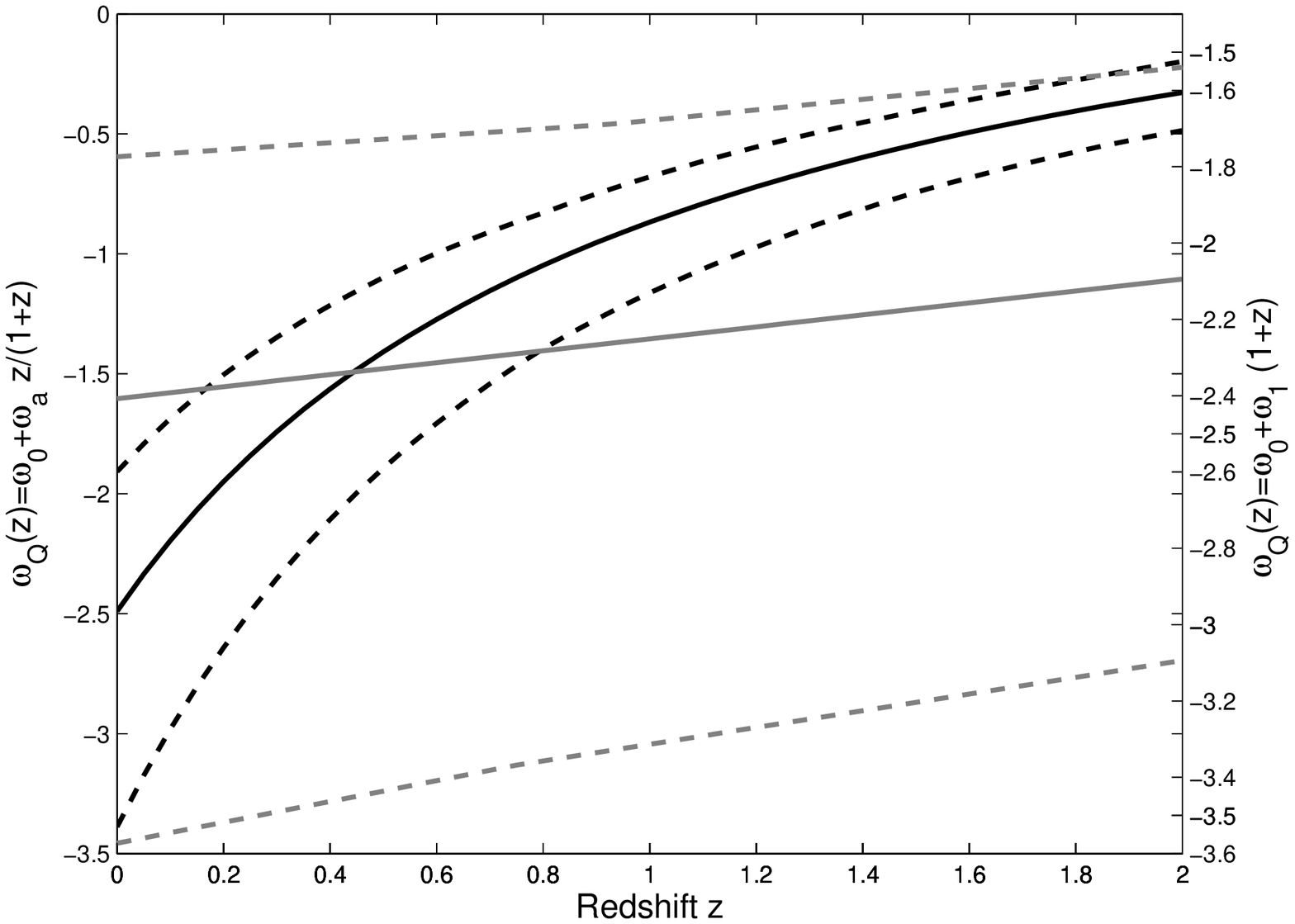}
\end{array}$ \end{center} \vspace{-0.2in} \caption{The best
supernova and WMAP data fits to the polynomial model and linear
model. The left panel shows Riess gold sample and WMAP data fits
to the two parameter polynomial model, the light black lines are
for $\Omega_{\rm Q}$ and the dark balck lines for $\omega_{\rm
Q}$, the solid lines are from the best fit. The right panel shows
Riess gold sample and WMAP data fits to the two parameter linear
model of $\omega_{\rm Q}$, the solid lines are from the best fit,
the light black lines are for the linear model and the dark black
lines are for the stable model. The dashed lines define the
1$\sigma$ boundaries.} \label{wqzfit} \end{figure}
\section{Model-independent Results} To construct a model
independent result, we first parameterize the dark energy density
by two parameters \cite{alam}, $\Omega_{\rm
Q}(z)=A_0+A_1(1+z)+A_2(1+z)^2$, here $\Omega_{\rm Q}(z)=8\pi
G\rho_{\rm Q}(z)/(3H^2_0)$ and $A_0=1-\Omega_{\rm m0}-\Omega_{\rm
r0}-A_1-A_2$. The relationship between the dark energy state of
equation parameter $\omega_{\rm Q}$ and the redshift is
$$\omega_{\rm Q}={1+z\over 3}{A_1+2A_2(1+z)\over
A_0+A_1(1+z)+A_2(1+z)^2}-1.$$ With the above parameteriaztion, we
find that $\Omega_{\rm Q}\ll \Omega_{m}$ and $\omega_{\rm
Q}\approx -1/3$ when $z\gg 1$. The best fit parameters to Riess
gold sample and WMAP data are $\Omega_{\rm m0}=0.40\pm 0.14$,
$A_1=-3.6^{+3.8}_{-4.6}$ and $A_2=1.23^{+1.74}_{-1.23}$ with
$\chi^2=174.3$. By using the best fit parameters, we find that
$\omega_{\rm Q0}=-1.4$ and $z_{\rm T}=0.37$. Then we consider the
commonly used two parameter linear model $\omega_{\rm
Q}(z)=\omega_0+\omega_1(1+z)$. The Riess gold sample and WMAP data
give that $\Omega_{\rm m0}=0.49^{+0.09}_{-0.20}$,
$\omega_0=-2.57^{+1.58}_{-5.18}$ and
$\omega_1=0.16^{+0.53}_{-0.16}$ with $\chi^2=173.7$. Combining the
best fit parameters, it is found that $\omega_{\rm Q0}=-2.41$ and
$z_{\rm T}=0.29$. Because the above linear model is divergent as
$z\gg 1$, so we next consider a more stable prarmeterization of
$\omega_{\rm Q}(z)=\omega_0+\omega_a z/(1+z)$ \cite{linder}. By
using this parameterization, we find that the best fit parameters
to Riess gold sample and WMAP data are $\Omega_{\rm
m0}=0.47^{+0.10}_{-0.19}$, $\omega_0=-2.5^{+1.5}_{-4.9}$ and
$\omega_a=3.2^{+6.6}_{-3.2}$ with $\chi^2=173.5$. Therefore the
turnaround redshift is $z_{\rm T}=0.28$ and $\omega_{\rm
Q0}=-2.5$. Note that the cosmological constant model $\omega_0=-1$
and $\omega_a=0$ is at the boundary of the $1\sigma$ parameter
space. However, the dark energy term became the dominant term when
$z\gg 1$ since $\omega_0+\omega_a=0.7$. Therefore, we use the
results from the polynomial parameterization only. The evolutions
of $\Omega_{\rm Q}(z)$ and $\omega_{\rm Q}(z)$ with redshift are
shown in figure \ref{wqzfit}.

For the model independent second order polynomial
parameterization, we find that $\Omega_{\rm m0}=0.40\pm 0.14$,
$\omega_{\rm Q0}=-1.4$ and $z_{\rm T}=0.37$. Alam, Sahni and
Starobinsky obtained $\Omega_{\rm m0}=0.385$ and $z_{\rm
T}=0.39\pm 0.03$ in a similar analysis \cite{alam}. Tegmark {\it
et al.} found that $\Omega_{\rm m0}\approx 0.30\pm 0.04$ by using
the WMAP data in combination with the Sloan Digital Sky Survey
(SDSS) data \cite{tegmark}. More recently, Riess {\it et al.}
showed that $z_{\rm T}=0.46\pm 0.13$ from the two parameter linear
model by using SNe Ia data only with the assumption that
$\Omega_{\rm m0}=0.27\pm 0.04$ \cite{riess}. The above results are
consistent with each other. For a flat universe with constant
$\omega_{\rm Q}$, we find that $\Omega_{\rm
m0}=0.30^{+0.06}_{-0.08}$ and $\omega_{\rm Q}\le -0.82$. The
result is consistent with our model independent results and that
in Refs.~\refcite{raknop03,riess,weller}. With those parameter
values, we find that the turnaround redshift $0.47\le z_{\rm T}\le
0.95$. For the constant pressure model $\beta=1/2$, the best fits
to the combined supernova and WMAP data are $\Omega_{\rm
m0}=0.298$ and $\omega_{\rm Q0}=-0.985$ which result in $z_{\rm
T}=0.65$. The best parameter fits to the combined supernova and
WMAP data for the general model analyzed in this paper are
$\Omega_{\rm m0}\sim 0.3$, $\omega_{\rm Q0}\sim -1.0$ and
$\beta\sim 0.5$. The turnaround redshfit is $z_{\rm T}\sim 0.67$.
These results are consistent with the observations. In conclusion,
it is shown that the general model in Ref.~\refcite{gong02} is
consistent with current observations and the model effectively
tends to be the $\Lambda$-CDM model. Furthermore, our model
independent results support the conclusion of dark energy
metamorphosis obtained in Ref. ~\refcite{alam}.

\section*{Acknowledgements}
The author thanks M. Doran for pointing our his original work in
the parameterization of the peaks of CMB power spectrum. The
author thanks D. Polarski for kindly pointing out his original
work on the stable parameterization. The author is grateful for
the anonymous referee's comments. This work is supported by CQUPT under grant
Nos. A2003-54 and A2004-05, NNSFC under grant No. 10447008 and CSTC under grant No. 2004BB8601.


\begin{thebibliography}{sp97}
\bibitem{sp97} S. Perlmutter {\it et al.},
{\it Nature} {\bf 391}, 51 (1998); {\it Astrophys. J.} {\bf 517},
565 (1999); P.M. Garnavich {\it et al.}, {\it Astrophys. J. Lett.}
{\bf 493}, L53 (1998); A.G. Riess {\it et al.}, {\it Astron. J.}
{\bf 116}, 1009 (1998).
\bibitem{tonry} J.L. Tonry {\it et al.}, {\it Astrophys. J.} {\bf 594}, 1
(2003); B.J. Barris {\it et al.}, {\it ibid.} {\bf 602}, 571
(2004)
\bibitem{raknop03} R.A. Knop {\it et al.}, {\it Astrophys. J.} {\bf 598}, 102
(2003);.
\bibitem{riess} A.G. Riess {\it et al.}, {\it Astrophys. J.} {\bf 607}, 665 (2004).
\bibitem{pdb00} P. de Bernardis {\it et al.}, {\it Nature} {\bf 404}, 955 (2000);
S. Hanany {\it et al.}, {\it Astrophys. J. Lett.} {\bf 545}, L5 (
2000).
\bibitem{bennett03} C.L. Bennett {\it et al.}, {\it Astrophys. J. Supp. Ser.}  {\bf
148}, 1 (2003).
\bibitem{DNSpergel} D.N. Spergel {\it et al.}, {\it Astrophys. J. Supp. Ser.} {\bf 148}, 175
(2003).
\bibitem{agr} A.G. Riess, {\it Astrophys. J.} {\bf
560}, 49 (2001).
\bibitem{mstagr} M.S. Turner and A.G. Riess, {\it Astrophys. J.} {\bf 569}, 18
(2002); R.A. Daly and S.G. Djorgovski, {\it ibid.} {\bf 597}, 9
(2003); {\bf 612}, 652 (2004).
\bibitem{holo} A. Cohen, D. Kaplan and A. Nelson, {\it Phys. Rev.
Lett.} {\bf 82}, 4971 (1999); S.D.H. Hsu, {\it Phys. Lett.} {\bf
B594}, 13 (2004); R. Horvat, {\it Phys. Rev.} {\bf D70}, 087301 (2004); M. Li,
{\it Phys. Lett.} {\bf B603}, 1 (2004); Q.G. Huang and Y. Gong, {\it J. Cosm. Astropart.
Phys.} {\bf 0408}, 006 (2004); Y. Gong, {\it Phys. Rev.} {\bf
D70}, 064029 (2004).
\bibitem{phantom} R.R. Caldwell, {\it Phys. Lett.} {\bf B545},
23 (2002); A. Melchiorri, I. Mersini, C.J. Odman and M. Trodden,
{\it Phys. Rev.} {\bf D68}, 043509 (2003); G.W. Gibbons,
hep-th/0302199; S.M. Carroll, M. Hoffman and M. Trodden, {\it
Phys. Rev.} {\bf D68}, 023509 (2003); J.G. Hao and X.Z. Li, {\it
Phys. Rev.} {\bf D67}, 107303 (2003); {\bf D70}, 043529 (2004);
{\it Phys. Lett.} {\bf B606}, 7 (2005); P. Singh, M. Sami and N. Dadhich, {\it Phys.
Rev.} {\bf D68}, 023522 (2003); J.S. Alcaniz, {\it Phys. Rev.}
{\bf D69}, 083521 (2004); M. Kaplinghat and S. Bridle,
astro-ph/0312430.
\bibitem{chaply} A. Kamenshchik, U. Moschella and V. Pasquier, {\it Phys. Lett.}
{\bf B511}, 265 (2001); N. Bilic, G. Tupper and R.D. Viollier,
{\it ibid.} {\bf B535}, 17 (2002); M.C. Bento, O. Bertolami and
A.A. Sen, {\it Phys. Rev.} {\bf D66}, 043507 (2002); D. Carturan
and F. Finelli, {\it ibid.} {\bf D68}, 103501 (2003); J.V. Cunha,
J.S. Alcaniz and J.A.S. Lima, {\it ibid.} {\bf D69}, 083501
(2004); L. Amendola, F. Finelli, C. Burigana and D. Carturan, {\it
J. Cosm. Astropart. Phys.} {\bf 0307}, 005 (2003).
\bibitem{quint} R.R. Caldwell, R. Dave and P.J. Steinhardt,
{\it Phys. Rev. Lett.} {\bf 80}, 1582 (1998); I. Zlatev, L. Wang
and P.J. Steinhardt, {\it ibid.} {\bf 82}, 896 (1999).
\bibitem{pgfmj} P.G. Ferreira and M. Joyce, {\it Phys. Rev. Lett.} {\bf
79}, 4740 (1997); P.G. Ferreira and M. Joyce, {\it Phys. Rev.}
{\bf D58}, 023503 (1998).
\bibitem{brpjep} B. Ratra and P.J.E. Peebles, {\it Phys. Rev.} {\bf D37}, 3406
(1988); C. Wetterich, {\it Nucl. Phys.} {\bf B302}, 668 (1988).
\bibitem{ptw99} S. Perlmutter, M.S. Turner and M. White, {\it Phys. Rev. Lett.} {\bf 83},
670 (1999).
\bibitem{johri} V. Sahni and A.A. Starobinsky, {\it Int. J. Mod. Phys.} {\bf D9}, 373
(2000); C. Rubano and J.D. Barrow, {\it Phys. Rev.} {\bf D64},
127301 (2001); V.B. Johri, {\it Class. Quantum Grav.} {\bf 19},
5959 (2001).
\bibitem{lautm} L.A. Ure\~{n}a-L\'{o}pez and T. Matos, {\it
Phys. Rev.} {\bf D62}, 081302 (2000); E. Di Pietro and J. Demaret,
{\it Int. J. Mod. Phys.} {\bf D10}, 231 (2001).
\bibitem{aasss} A.A. Sen and S. Sethi, {\it Phys. Lett.} {\bf
B532}, 159 (2002).
\bibitem{gong02} Y. Gong, {\it Class. Quantum Grav.} {\bf 19}, 4537
(2002).
\bibitem{tachyon} C. Armendariz-Picon, T. Damour and V. Mukhanov, {\it Phys. Lett.} {\bf B458}, 209
(1999); T. Padmanabhan and T.R. Choudhury, {\it Phys. Rev.} {\bf
D66}, 081301 (2002); J.S. Bagla, H.K. Jassal and T. Padmanabhan,
{\it ibid.} {\bf D67}, 063504 (2003); T. Padmanabhan, {\it Phys.
Rep.} {\bf 380}, 235 (2003); T. Padmanabhan and T.R. Choudhury,
{\it Mon. Not. Roy. Astron. Soc.} {\bf 344}, 823 (2003).
\bibitem{freese02} K. Freese and M. Lewis, {\it Phys.Lett.} {\bf B540}, 1
(2002); K. Freese, {\it Nucl. Phys. Suppl.} {\bf 124}, 50 (2003).
\bibitem{gondolo} P. Gondolo and K. Freese, {\it Phys. Rev.} {\bf D68}, 063509
(2003).
\bibitem{sen03} S. Sen and A.A. Sen, {\it Astrophys. J.} {\bf 588}, 1 (2003)
; A. A. Sen and S. Sen, {\it Phys. Rev.} {\bf D68}, 023513 (2003).
\bibitem{zhu03} Z.H. Zhu and M. Fujimoto, {\it Astrophys. J.} {\bf 581}, 1
(2002); {\bf 585}, 52 (2003); {\bf 602}, 12 (2004); Z.H. Zhu, M.
Fujimoto and X.T. He, {\it {ibid.}} {\bf 603}, 365 (2004).
\bibitem{wang} Y. Wang, K. Freese, P. Gondolo and M. Lewis, {\it Astrophys. J.} {\bf 594}, 25
(2003).
\bibitem{maltamaki} T. Multamaki, E. Gaztanaga and M. Manera, {\it Mon. Not. Roy. Astron. Soc.}
{\bf 344}, 761 (2003).
\bibitem{frith} W.J. Frith, {\it Mon. Not. Roy. Astron. Soc.}
{\bf 348}, 916  (2004).
\bibitem{gong03} Y. Gong and C.K. Duan,
{\it Class. Quantum Grav.} {\bf 21}, 3655 (2004); {\it Mon. Not.
Roy. Astron. Soc.} {\bf 352}, 847 (2004); Y. Gong, X.M. Chen and
C.K. Duan, {\it Mod. Phys. Lett.} {\bf A19}, 1933 (2004).
\bibitem{doran} W. Hu, M. Fukugita, M. Zaldarriaga and M. Tegmark, {\it Astrophys. J.} {\bf
549}, 669 (2001); M. Doran, M. Lilley, J. Schwindt and C.
Wetterich, {\it ibid.} {\bf 559}, 501 (2001); M. Doran, M. Lilley
and C. Wetterich, {\it Phys. Lett.} {\bf B528}, 175 (2002); M.
Doran and M. Lilley, {\it Mon. Not. Roy. Astron. Soc.} {\bf 330},
965 (2002).
\bibitem{wang1} J.R. Bond, G. Efstathiou and M. Tegmark, {\it Mon. Not. Roy. Astron. Soc.} {\bf
291}, L33 (1997); A. Melchiorri, L. Mersini, C.J. \"{O}dman and M.
Trodden, {\it Phys. Rev.} {\bf D68}, 043509 (2003); Y. Wang and P.
Mukherjee, {\it Astrophys. J.} {\bf 606}, 654 (2004); Y. Wang and
M. Tegmark, {\it Phys. Rev. Lett.} {\bf 92}, 241302 (2004).
\bibitem{alam} U. Alam, V. Sahni, T.D. Saini and A.A. Starobinsky,
{\it Mon. Not. Roy. Astron. Soc.} {\bf 354}, 275 (2004);
U. Alam, V. Sahni and A.A. Starobinsky, {\it J.
Cosm. Astropart. Phys.} {\bf 0406}, 008 (2004); Y. Gong, Class. Quantum Grav., in press,
astro-ph/0405446.
\bibitem{linder} M. Chevallier and D. Polarski, {\it Int. J. Mode. Phys.} {\bf D10}, 213 (2001);
E.V. Linder, {\it Phys. Rev. Lett.} {\bf 90},
91301 (2003).
\bibitem{tegmark} M. Tegmark {\it et al.},
{\it Phys. Rev.} {\bf D69}, 103501 (2004).
\bibitem{weller} J. Weller and
A.M. Lewis, {\it Mon. Not. Roy. Astron. Soc.} {\bf 346}, 987
(2003); P. Schuecker {\it et al.}, {\it Astron. Astrophys.} {\bf
402}, 53 (2003).
\end{thebibliography}
\end{document}